\documentclass[aps,prb,preprint, 
superscriptaddress,balancelastpage,nobibnotes]{revtex4-1}
\usepackage{amsmath}
\usepackage{amssymb}
\usepackage{graphicx}
\usepackage{float}
\usepackage{bm}
\usepackage{color}
\usepackage{natbib}
\usepackage{easyReview}
\usepackage{physics}
\setcitestyle{numbers}
\setcitestyle{square}
\usepackage[colorlinks=true,
urlcolor=blue,
linkcolor=blue,
citecolor=blue]{hyperref}
\usepackage{url}
\usepackage{booktabs}

\usepackage{multirow}
\usepackage{array}
\usepackage{tabularx}
\graphicspath{{../figures/}}

\newcommand{\kt}[1]{\left \lvert #1 \right \rangle}
\newcommand{\br}[1]{\left \langle #1 \right \rvert}
\makeatletter
\newcommand*{\rom}[1]{\expandafter\@slowromancap\romannumeral #1@}
\makeatother

\begin{document}
	
	\title{Sunburst quantum Ising battery under periodic delta-kick charging}
		\author{Ankita Mazumdar}
		\author{Akash Mitra}
		\author{Shashi C. L. Srivastava}
		\email{Corresponding author : shashi@vecc.gov.in}
		\affiliation{Variable Energy Cyclotron Centre, 1/AF Bidhannagar, Kolkata 
				700064, India}
		\affiliation{Homi Bhabha National Institute, Training School Complex, 
				Anushaktinagar, Mumbai - 400094, India}
	\begin{abstract}

   Most quantum batteries studied so far with notable exception of 
   Sachdev-Ye-Kitaev (SYK) 
   batteries  are based on integrable models, where superlinear 
   scaling of charging power—and hence a quantum advantage—can be achieved, but 
   at the cost of unstable stored energy due to integrability. Here, by 
   considering the sunburst quantum Ising battery driven by periodic 
   delta-kicks, we show that in the quantum 
   chaotic regime a 
   quantum advantage is achieved for number of batteries $n_b \leq 4$, together 
   with excellent stability of energy storage. In the integrable regime optimal 
   energy storage and extraction are possible irrespective of the initial state 
   of the charger. Finally, we show that the 
   observed advantage does not 
   originate from multipartite entanglement within the battery subsystem and is 
   therefore classical in nature.
		
	\end{abstract}
	\maketitle

	\section{Introduction}
	
	In recent times, there has been {an} upsurge of interest in 
	the field of quantum thermodynamics, where fundamental thermodynamic 
	principles such as energy transfer, storage, and work extraction processes 
	are investigated at the quantum level 
	\cite{Vinjanampathy_2016,binder_2019,gemmer_2009, Goold_2016, Bera_2019}. 
	The primary motivation has been to explore the role of quantum resources, 
	such as coherence and entanglement, in enhancing the performance of 
	different quantum thermodynamic devices, including quantum heat engines 
	\cite{Kosloff_2014, Scully_2003, Zhang_2007, Wang_2009,Hammam_2021, 
	Myers_2022} and quantum batteries (QBs) 
	\cite{Alicki_2013,Hovhannisyan_2013, Binder_2015, 
	Campaioli_2017,Andolina_2019,andolina_qm_class, Shi_2022, 
	MitraSrivastava_2024,Ferraro_2018,Le_2018,Dou_2022,Ghosh_2020,Ghosh_2021,Peng_2021,Liu_2021,Yang_2025,Rossini_2020,Francica_2024,Rosa_2020}.
	 Since the introduction of QBs by Alicki and Fannes \cite{Alicki_2013}, the 
	possibility of realizing faster charging by utilizing the quantum 
	entanglement has been investigated in details \cite{Hovhannisyan_2013, 
	Binder_2015, Campaioli_2017}.	
	The global entangling operations, which enable the system to traverse a 
	correlated shortcut in the associated Hilbert space causes the speed up in 
	charging process \cite{Hovhannisyan_2013, Binder_2015}. As a result, 
	collective charging via global entangling unitary operations leads to a 
	super-extensive scaling in charging power, known as the quantum advantage 
	\cite{Campaioli_2024}. The first realization of quantum advantage in 
	physically realizable models was shown in the Dicke QB \cite{Ferraro_2018}, 
	where the charging power per unit QB scales as $\sqrt{N}$, with $N$ being 
	the total number of QBs. The super-extensive scaling of charging power has 
	also been observed in other models, including spin chain batteries 
	\cite{Le_2018,Dou_2022,Ghosh_2020,Ghosh_2021,puri_2024}, central spin 
	batteries \cite{Peng_2021,Liu_2021,Yang_2025} and Sachdev-Ye-Kitaev 
	batteries \cite{Rossini_2020,Francica_2024}. Recently, $\sqrt{N_c}$ quantum 
	advantage in charging process with $N_c$ number of chargers and single 
	battery of the central spin QB has been experimentally verified in NMR 
	systems \cite{Joshi_2022}.

    Despite the presence of quantum advantage in the charging process, 
    simultaneous optimization of extracted energy and charging power was not 
    possible \cite{Ferraro_2018,Le_2018,Andolina_2018}. In the recently 
    proposed sunburst quantum Ising battery \cite{MitraSrivastava_2024}, such 
    simultaneous optimization became possible, together with the independence 
    of both ergotropy—defined as the maximum amount of extracted energy via 
    unitary 
    operations—and charging power from the choice of the initial state. 
    However, the charging power in this model was shown to scale only linearly 
    with the number of batteries, indicating the absence of a quantum 
    advantage. Although integrable models have been shown to provide quantum 
    advantage in 
    charging \cite{Yang_2025,Grazi,thao_p_le}, they are unlikely to guarantee 
    stability due to finite memory effects of the initial state, which induce 
    rapid fluctuations in stored energy. It is also desirable for a QB not only 
    to store the optimal amount of energy but also to allow complete extraction 
    of the stored energy through unitary operations. In the case of a central 
    spin QB, such optimization can be realized by tuning the initial state of 
    the charger, although this remains an experimentally challenging task 
    \cite{Yang_2025}.
	
	The effect of periodically driven charging process of a QB, modelled by 
	transverse Ising 
	chain, with both transverse and longitudinal magnetic field, has been 
	explored in Ref.~\cite{Mondal_2022}. Although global charging has been
	achieved in that setup, no quantum advantage in charging power was 
	observed. The periodically driven charging protocol 
	was further 
	explored in Ref.~\cite{puri_2024}, where it was shown that in the presence 
	of both long-range interactions and a periodically driven charging, a 
	superlinear scaling of charging power can be achieved in the long range 
	$XY$ model. There is a clear gap in current understanding of whether 
	one can arrange optimal storage, quantum advantage, stability in the stored 
	energy in a single model. Also, whether long range interactions are 
	absolutely needed along with periodic driving to get the quantum advantage 
	as has been the case in Ref.~\cite{puri_2024}? To address these, we 
	investigate the sunburst quantum Ising battery \cite{MitraSrivastava_2024} 
	which involves only nearest neighbor interactions driven by periodic 
	delta-kicks. We 
	analyze 
	the optimal energy storage and work extraction in this model of quantum 
	battery numerically with some analytical results in limiting cases. We also 
	demonstrate the stability of the energy storage and quantum advantage in 
	the same model. Our results provide a unique situation where not only we 
	achieve all the key markers of the quantum battery in the same model, we 
	also demonstrate that quantum advantage can be arranged via periodic 
	driving in short-range interacting models as well.

	The paper is organized as follows. In Sec.~\ref{sec:rmt}, we study the 
	statistical properties of the model under periodic delta-kicks. In 
	Sec.~\ref{sec:optimal_energy_storage}, we study 
	different limiting cases of the model where it is possible to realize 
	optimal energy storage and work extraction. We show the stability of the 
	energy storage and quantum advantage in {periodic }charging process 
	in 
	Sec.~\ref{sec:stability_advantage}, and discuss the nature of advantage in 
	Sec.~\ref{sec:adv_nature}, while we present a conclusion in 
	Sec.~\ref{sec:conc}.
	
	\section{Statistical properties of kicked sunburst quantum Ising model}\label{sec:rmt}
	
	The sunburst quantum Ising model, which consists of a transverse field 
	Ising chain symmetrically coupled to a few external isolated qubits, has 
	been studied in the context of analyzing ground state properties 
	\cite{Franchi_2022} and entanglement dynamics under interaction quench 
	protocol \cite{Mitra_2024}. Recently, this model has been explored as a 
	quantum battery by considering the transverse Ising chain as a charger and 
	external qubits as a quantum battery, known as the sunburst quantum Ising 
	battery \cite{MitraSrivastava_2024}. The Hamiltonian of this charger and 
	battery is
	\begin{equation}\label{eq:tot_ham}
		H = H_c \otimes \mathbb{I}_{b} + \mathbb{I}_{c} \otimes H_b + \lambda(t) V_{cb},
	\end{equation}
	where $H_c$ and $H_b$ represent the Hamiltonian of the charger and battery, 
	respectively. The interaction Hamiltonian is represented by $V_{cb}$, while 
	$\mathbb{I}_{b}$ ($\mathbb{I}_{c}$) represents the identity operator in the 
	space of the battery (charger). Here, $\lambda(t)$ represents a 
	time-dependent external control parameter, {defining the charging 
	protocol. }
	 The Hamiltonian for charger and the battery along with the interaction 
	 are expressed as
	\begin{equation}
		\begin{aligned}
			H_c&=-\sum_{i=1}^L \left(J\sigma_i^x \sigma_{i+1}^x 
			+h\sigma_i^z\right)\\
			H_b &= -\frac{\delta}{2}\sum_{i=1}^{n_b} \Sigma_i^z; 
			\quad 
			{\color{black}	V_{cb} = -\kappa \sum_{i=1}^{n_b} \sigma_{1+(i-1)d}^x 
				\Sigma_{i}^x},
		\end{aligned}
	\end{equation}
	where $L$ and $n_b$ denote the number of Ising sites and external qubits, 
	respectively. $J$ is the strength of nearest neighbor interaction, which is 
	assumed to be positive, i.e., $J>0$ to ensure ferromagnetic interaction 
	between the spins, and $h$ is the transverse field strength. Furthermore, 
	$\sigma_i$ denotes the Pauli matrix on the $i$th Ising site, while 
	$\Sigma_i$ denotes the Pauli matrix {corresponding to } the 
	$i$th battery. The energy gap between the two lowest eigenstates of the 
	battery is represented as $\delta$ and $\kappa$ represents the interaction 
	strength between the battery and the charger. The distance between two 
	consecutive qubits is denoted as $d$.

	In this paper, we { focus on the periodically driven charging 
	protocol}, specifically of the form
	\begin{equation}
		\lambda(t)=\sum_{n=-\infty}^{\infty}\delta \left( \frac{t}{\tau} 
		-n\right),
	\end{equation}
	where $\tau$ denotes the time period of charging Hamiltonian such that $\lambda(t+\tau)=\lambda(t)$. Such $\delta$-kicked quench protocol can be realized with ultracold atoms exposed to a pulsed standing wave of near-resonant light \cite{Moore_1995}. It is worth noting that the above quench protocol differs from refs.~\cite{Mondal_2022,puri_2024} where the periodic driving has a square wave form instead of a delta function impulse. 
	
	Given the periodic structure of the Hamiltonian in Eq.~\ref{eq:tot_ham}, 
	i.e., $H(t+\tau)=H(t)$, the unitary operator $U$ {for one 
	time-period} describing the time evolution from just before to just after 
	the 
	kick can be expressed in a factorized form as
	\begin{align}\label{eq:Unitaryop_generic}
		U &= e^{-i V_{cb} \tau}e^{-i \left(H_c \otimes \mathbb{I}_{b} + 
		\mathbb{I}_{c} \otimes H_b\right) \tau} \\
		&=e^{-i V_{cb} \tau}\left[e^{-iH_c \tau} \otimes e^{-iH_b \tau} 
		\right]\equiv U_{cb}\left(U_c\otimes U_b\right) \,  ,
	\end{align}
	where $U_c (U_b)$ governs the time evolution {of charger (battery) 
	alone in absence of the interaction} {while $U_{cb}$} 
	accounts for their instantaneous coupling via the interaction $V_{cb}$.
	
	To understand the statistical properties of {the quasi-energies $\phi_m$ 
	and eigenvectors $\kt{\psi_m}$, defined 
	by, }, 
	\begin{equation}\label{eq:Ueigvals}
		U  |\psi_m\rangle= e^{i \phi_m} |\psi_m\rangle.
	\end{equation}
	of this Floquet system, we take resort to random matrix theory.
	The spacing distribution of consecutive energy levels is a widely used tool 
	to distinguish integrable systems from classically chaotic ones. In chaotic 
	systems, the level spacings typically follow the Wigner-Dyson distribution, 
	whereas integrable systems exhibit a Poisson distribution 
	\cite{Bohigas_1984,Haake1991,Berry_1977}. The spacing between two 
	consecutive {unfolded quasi-energies} is defined as
	\begin{equation}\label{eq:spacings}
		s_m=\frac{D}{2 \pi}\left(\phi_{m+1}-\phi_m\right),
	\end{equation}
	where $D$ the dimensionality of the Hilbert space \cite{Arnd_2023}. The prefactor in the above equation ensures the unit mean spacing. If a system exhibits quantum chaos, both its eigenstates and eigenphases are expected to satisfy the predictions of random matrix theory (RMT). According to RMT, in the presence of time-reversal symmetry (Circular Orthogonal Ensemble ), the nearest-neighbor level spacing distribution follows the Wigner-Dyson form:
	\begin{equation}
		P(s)=\frac{\pi}{2} s \exp \left(-\frac{\pi}{4} s^2\right).
	\end{equation}
{The ratio of nearest neighbor spacing, $\Tilde{r}_n$, }
	another widely used quantity in RMT for its independence on unfolding 
	scheme to study spectral fluctuation properties {is }defined as 
	\cite{Huse_2007}
	\begin{equation}\label{eq:avg_ratiospacings}
		\Tilde{r}_n=\frac{\min (s_n,s_{n-1})}{\max(s_n,s_{n-1})}.
	\end{equation}
	The average value of the ratio of nearest neighbor spacing, $\langle 
	\Tilde{r}\rangle$, for the circular orthogonal ensemble takes the value 
	$\langle \Tilde{r}\rangle \approx 0.53$ while for integrable systems 
	$\langle \Tilde{r}\rangle \approx 0.38$ \cite{Atas_2013}. 
    
	
	\begin{figure}[H]
		\centering
		\includegraphics{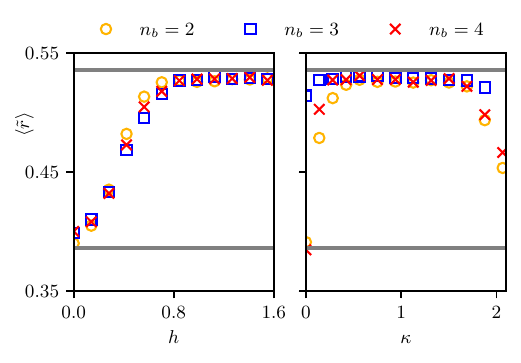}
		\caption{(Left) The variation of $\langle \tilde{r} \rangle $ with $h$ 
		is plotted by keeping $J=\delta = \kappa=1$ for three different number 
		of batteries, $n_b=2,3,4$. A clear transition from integrability for 
		$h= 0$ to quantum chaotic 
		nature for $h\approx 1$  is 
		captured through average ratio of spacing. Two solid gray lines 
		correspond to $\langle \tilde{r}\rangle  \approx 0.386$ and  $\langle 
		\tilde{r}\rangle  \approx 0.53$. (Right) 
		Same is plotted  
		with $\kappa$ by keeping $J=\delta = h=1$ for three different number 
		of batteries, $n_b=2,3,4$. The expected transition from integrability 
		to chaos with increasing $\kappa$ is clearly seen. The pronounced dip 
		at $\kappa=2$ is due to $U_{cb}$ becoming a local operator (Eq. 
		\ref{eq:U_specialcase}) which maps a direct product state to another 
		direct product state.}
		\label{fig:rmt}
	\end{figure}
    Let's note that statistical features of a system are greatly affected by 
    the presence of symmetries, and therefore, before comparing with RMT, one 
    needs to either break all the discrete symmetries or compare the results 
    for symmetry reduced operator. The translational symmetry and qubit 
    exchange symmetry  of 
    the Floquet system are broken 
    by randomizing the external magnetic field strength, $h$, and $\delta$ near 
    the mean value. Since the Hamiltonian commutes with the parity 
    operator, we restrict the unitary $U$ to the even‑parity sector.
	As we will see in the subsequent 
    section, identifying the nature of the Hamiltonian $H$ in 
    Eq.~\ref{eq:tot_ham}—whether integrable or quantum chaotic—plays a pivotal 
    role in understanding the performance of the QB {and as the 
	average values of ratio of spacings itself characterize the degree of chaos 
	in the system satisfactorily, we plot $\langle \Tilde{r}_n\rangle$ as a 
	function of $h$ and interaction strength $\kappa$ in Fig.} \ref{fig:rmt}, 
	by 
    fixing $(\kappa, \delta)$ and $(h/J\color{black}, \delta)$ respectively. In 
    the limit $\kappa \to 0$, the system is in the integrable regime-this is 
    expected since both subsystems, namely the transverse field Ising model and 
    the external isolated qubits, are integrable. As previously reported in the 
    case of the sunburst quantum Ising model with $\lambda(t)=1$ 
    \cite{Mitra_2024}, the total system approaches the near integrable regime 
    in the limit $h/J \ll 1$. However, with increasing 
    $\kappa$ and for $h/J \sim 1$, the system is in the quantum chaotic regime. 
    Unlike the static case $\lambda(t)=1$, under periodic delta-function 
    impulses, the total system undergoes a rapid transition to the quantum 
    chaotic regime even at relatively small interaction strengths, such as 
    $\kappa \sim 0.2$. It is also worth noting that for $\kappa \tau=\pi/2$, 
    $\langle \Tilde{r}\rangle$ exhibits a sharp dip. This behavior can be 
    understood from the structure of the unitary operator $U_{cb}$, which in 
    this limit takes the form
	\begin{equation}\label{eq:U_specialcase}
		U_{cb}= e^{i \frac{\pi}{2}\left(\sigma_{1+(i-1)d}^x \otimes
			\Sigma_{i}^x\right)}=i\prod_{i=1}^{n}  \sigma_{1+(i-1)d}^x \otimes
		\Sigma_{i}^x.
	\end{equation}
	It is evident from the above expression that $U_{cb}$ simply flips all the 
	Ising and qubit spins along the $z$ direction. Any arbitrary state in the 
	Hilbert space of the total system can be expressed as 
	\begin{equation}
		|\Phi\rangle=\sum_{i=1}^{2^L} \sum_{j=1}^{2^{n_b}} \beta_{ij} 
		|c_i\rangle  \otimes |b_j\rangle,
	\end{equation}
	where $|c_i\rangle \, ,|b_j\rangle$ denote the $i$-th eigenstate of the transverse field Ising chain and the $j$-th eigenstate of external isolated qubits, respectively. It is easy to verify that if $|\Phi\rangle$ is an eigenstate of the unitary operator $U$ in Eq.~\ref{eq:Unitaryop_generic}, then the coeffients $\beta_{ij}$ must be identical for all $j$. This allows us to simplify the generic state  $|\Phi\rangle$ to the form
	\begin{equation}
		|\Phi\rangle=\sum_{i=1}^{2^L} \beta_i |c_i\rangle \otimes 
		\sum_{j=1}^{2^{n_b}} |b_j\rangle.
	\end{equation}
	This shows that $U_{cb}$ in eq.~\ref{eq:U_specialcase} acts as a local 
	operator and fails to entangle the two subsystems.
	The above form of $|\Phi\rangle$ indicates that, for $\kappa \tau=\pi/2$, 
	the eigenstates of $U$ can be written as direct product states between the 
	two individual subsystems. This, in turn, implies that for this specific 
	value of $\kappa$, the total system effectively behaves as two decoupled 
	non-interacting subsystems. Since both the individual subsystems are 
	integrable, it is natural to expect that the total system exhibits 
	integrable like behavior at $\kappa \tau=\pi/2$, {which} 
	explains the observed dip in the average level spacing ratio $\langle 
	\Tilde{r} \rangle$ at $\kappa \tau = \pi/2$ as shown in 
	Fig.~\ref{fig:rmt}.

	\section{Optimal energy storage and work extraction}\label{sec:optimal_energy_storage}
	Unlike most QB models that lie within the integrable regime, the sunburst 
	quantum Ising battery subjected to periodic delta-{kicks} falls in the 
	quantum chaotic regime (see 
	Fig.~[\ref{fig:rmt}]). This presents a significant challenge for obtaining 
	an analytical insight about the performance of this QB. However, as 
	mentioned earlier, in the limit $h/J \ll 1$, the system makes a transition 
	to the near integrable regime. This allows us to analytically evaluate 
	various figures of merit related to the charging behavior of the QB and 
	work extraction. 
	
	We consider the initial state of the total system as a direct product state between the individual ground states of the two subsystems. For $h =0$, the ground state of the transverse field Ising model is the well-known “cat state” \cite{Sch1935}, also referred to as the Greenberger-Horne-Zeilinger state \cite{GHZ2007}, given by $ \kt{\psi^I_G} = 
	\frac{1}{\sqrt{2}}[ \kt{+++\dots+} + \kt{---\dots-}$ with $\sigma_x \kt{\pm} = \pm \kt{\pm}$. We focus on the case of a single battery, so the initial state of the total system is $\kt{\psi(0^-)}= \kt{\psi^I_G} \otimes \kt{0}$, where $\Sigma_z \kt{0} = \kt{0}$. Acting with the interaction operator $V_{cb}$ on $\kt{\psi^I_G}$ yields a new state $ \kt{\psi^I_N} = 
	\frac{1}{\sqrt{2}}[ \kt{+++\dots+} - \kt{---\dots-}$, which is orthogonal to $\kt{\psi^I_G}$. Applying $V_{cb}$ once more on $\kt{\psi^I_N}$ returns the system to its original state $\kt{\psi^I_G}$. Both the states $\kt{\psi^I_N}$ and $\kt{\psi^I_G}$ are eigenstates of the Ising Hamiltonian $H_c$ with same energy eigenvalue $-J$. As a result, $e^{-i \left( H_c \otimes \mathbb{I}_{b}\right) T}$ only produces a global phase factor and its effect can be ignored. This indicates that, in the limit $h \to 0$ and with only a single qubit, the dynamics produced by the unitary operator $U$ becomes independent of the number of Ising sites. Therefore, we can perform our analytical calculation by considering the simplest situation where the number of Ising sites is only 2, i.e., $L=2$. Moreover, the Hamiltonian commutes with the total parity operator, i.e, $[H,P]=0$, where $P=\displaystyle{\prod_{i=1}^L \sigma_i^z}\otimes\displaystyle{\prod_{j=1}^{n_b} \Sigma_j^z}$. This allows us to restrict ourselves in either even sector ($P=+1$) or odd parity sector ($P=-1$). With these considerations, we choose the basis states as
	\begin{align}
		\begin{cases}
			\kt{\phi_0}&=\frac{1}{\sqrt{2}}[\kt{++}+ \kt{--}]\otimes \kt{0}\\
			\kt{\phi_1}&=\frac{1}{\sqrt{2}}[\kt{++}- \kt{--}]\otimes \kt{1}\\
			\kt{\phi_2}&=\frac{1}{\sqrt{2}}[\kt{+-}+ \kt{-+}]\otimes \kt{0}\\
			\kt{\phi_3}&=\frac{1}{\sqrt{2}}[\kt{+-}- \kt{-+}]\otimes \kt{1}
		\end{cases}
	\end{align}
	In this basis, the matrix corresponding to the unitary operator $U$ becomes
	\begin{align}
		U=  
		\begin{bmatrix}
			\cos{\kappa \tau}e^{i\epsilon} & i\sin{\kappa \tau}e^{-i\epsilon} 
			&0 & 0\\
			i\sin{\kappa \tau}e^{i\epsilon} & \cos{\kappa \tau}e^{-i\epsilon} & 
			0 & 0\\
			0 & 0 & \cos{\kappa \tau}e^{i\epsilon} & i\sin{\kappa 
			\tau}e^{-i\epsilon}\\
			0 & 0 & i\sin{\kappa \tau}e^{i\epsilon} & \cos{\kappa \tau} 
			e^{-i\epsilon}
		\end{bmatrix},
	\end{align}
	where $\epsilon=\frac{\delta \tau}{2}$.
	The state $|\psi(n)\rangle$ after applying the $n$-th kick over the initial state $\kt{\psi(0^-)}$ can be easily obtained as $|\psi(n)\rangle=F_n \kt{\psi(0^-)}$, where the matrix $F_n$ is defined as $F_n=U^n$. Since the total system is always in a pure state, the density matrix of the full system becomes $\rho_{cb}^n=|\psi(n)\rangle\langle \psi(n)|$. By taking a partial trace over the charger degree of freedom, we obtain the reduced density matrix of the battery as
	\begin{align}\label{eq:rdm_qb_nkicks}
		\rho_b^n&=\lvert (F_n)_{11}\rvert^2 \kt{0}\br{0}+ \lvert (F_n)_{21}\rvert^2 \kt{1}\br{1}\\
		&=\left(1-\lvert (F_n)_{21}\rvert^2 \right) \kt{0}\br{0}+ \lvert (F_n)_{21}\rvert^2 \kt{1}\br{1},
	\end{align}
	where $\sigma_z|1\rangle=-|1\rangle$ and
	\begin{equation*}
		{(F_n)}_{21}= \frac{\left(\left(e^{-i\epsilon}(a + b)\right)^n - 
		\left(e^{-i\epsilon}(a - b)\right)^n\right)}{2^n b} ie^{2id} 
		\sin{\kappa\tau},
	\end{equation*}
	with, $a= 1+ e^{2i\epsilon}\cos{\kappa\tau},b= \sqrt{a^2-4e^{2i\epsilon}}$

\begin{figure}[H]
		\centering
		\includegraphics{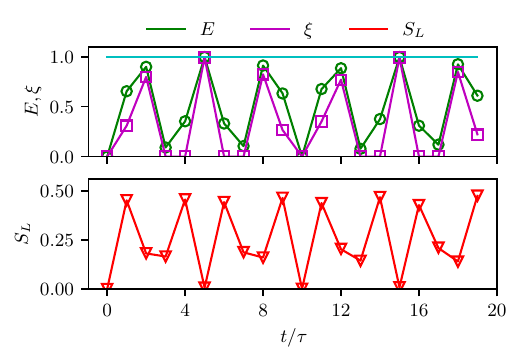}
		\caption{The dynamics for $E$(green), $\xi$ and $S_L$ are plotted for 
		the 
		system size $L=6$, $n_b=1$ by keeping the parameters $J,\delta=1$, 
		$\kappa=6$, $\tau= \pi/20$ and $h=0.1$. The symbols stand for the 
		numerical results and the solid lines are to show the analytical 
		results derived in Eqs. \ref{eq:stored_eng_expression}, 
		\ref{eq:ergotropy_expression} and \ref{eq:entropy_expression}. The 
		horizontal cyan line correspond to maximum possible stored energy, 
		$\delta$. }
		\label{fig:energy_storage}
	\end{figure}

	We now define a set of quantities that can be derived from the above 
	expression of the reduced density matrix of the battery and will be 
	necessary to analyze the performance of QB. The energy stored in the 
	battery after applying $n$ successive delta-kick is defined as
	\begin{equation}\label{eq:stored_eng_defn}
		E(n)=\tr \left(\rho_b^n H_b\right)-\tr \left(\rho_b^0 H_b\right),
	\end{equation}
	where $\rho_b^0$ is the density matrix of the battery corresponding to the initial state. Since we consider the initial state as the ground state of QB, we have $\tr \left(\rho_b^0 H_B\right)=-\delta/2$. Thus, following Eq.~\ref{eq:rdm_qb_nkicks} and Eq.~\ref{eq:stored_eng_defn}, the expression for stored energy becomes
	\begin{equation}\label{eq:stored_eng_expression}
		E(n)=\delta \rvert {(F_n)}_{21} \rvert^2.
	\end{equation}
	Since $\rvert {(F_n)}_{21} \rvert^2_{\max}=1$, the maximum amount of energy 
	that 
	can be stored in the QB is $\delta$, the energy gap between the ground 
	state and excited state of the battery. This is referred to as the optimal 
	energy storage. The another {figure of merit} of QB is the average charging 
	power which 
	quantifies the rate of charging in the QB and is defined at each 
	stroboscopic time $t=n \tau$ as
	\begin{equation}\label{eq:power_defn}
		P(n \tau)=  E(n \tau)/n \tau.
	\end{equation} 
	The charging time is defined as the time when stored energy is maximized- 
	if this happens after applying $m$ successive kicks, then the charging time 
	is simply $T=m \tau$. Under unitary cyclic transformation, not all the 
	stored amount of energy can be extracted. The maximum amount of useful 
	energy that can be extracted from the state $\rho_b^n$ via unitary 
	transformation $\Tilde{U}$ is known as the ergotropy 
	\cite{Allahverdyan_2004}, {and is }defined as
	\begin{equation}
		\xi(\rho_b^n)=\tr \left(\rho_b^n H_b\right)-\underset{\tilde{U}}{\min} 
		\left\{\tr \left(\Tilde{U}\rho_b^n \Tilde{U}^\dagger H_b\right)\right\},
	\end{equation}
	where the minimization has to be performed over all the unitary 
	transformations. It can be shown that the minimization is achieved when the 
	state after the unitary transformation coincides with its passive 
	counterpart \cite{Pusz_1978}, since any passive state $\tilde{\rho}_b^n$ 
	satisfies the condition $\tr \left(H_b \tilde{\rho}_b^n\right) \leq \tr 
	\left(H_b \tilde{U} \tilde{\rho}_b^n \tilde{U}^\dagger \right)$ for all 
	unitaries $\tilde{U}$. As a result, no energy can be extracted from the 
	passive state. The passive state $\tilde{\rho}_b^n$ must be diagonal in the 
	eigenbasis of the Hamiltonian and its eigenvalues {are the 
	eigenvalues of $\rho_b^n$, but only } in decreasing 
	order. With these considerations, the ergotropy is expressed as
	\begin{equation}\label{eq:erg_defn}
		\xi(\rho_b^n)=\tr \left(\rho_b^n H_b\right)-\tr \left(\tilde{\rho}_b^n 
		H_b\right).
	\end{equation}
	It is easy to see that for a single battery unless the occupation probability corresponding to the ground state $|0\rangle$ does not exceed the occupation probability corresponding to the excited state $|1\rangle$, ergotropy is zero since the passive state counterpart of $\rho_b^n$ coincides with the state $\rho_b^n$. The ergotropy is non-zero only when 
	\begin{equation}
		\lvert {(F_n)}_{11}\rvert^2 < \lvert {(F_n)}_{21}\rvert^2 \implies 
		\lvert {(F_n)}_{21}\rvert^2 >\frac{1}{2},
	\end{equation}
	or in other words population inversion occurs. If the above condition is satisfied, following Eq.~\ref{eq:rdm_qb_nkicks}, the expression of nonzero ergotropy is obtained as
	\begin{equation}\label{eq:ergotropy_expression}
		\xi(\rho_b^n)=\delta \left(2\lvert {(F_n)}_{21}\rvert^2 -1\right).
	\end{equation}
	Notice that optimal energy extraction—i.e., when the extracted energy 
	equals the energy gap between the ground and the most excited state, 
	$W(\rho_b^n)=\delta$ is possible only if $\lvert F_{21}\rvert^2=1$. This 
	condition implies that the occupation probability of the excited state 
	becomes unity, while that of the ground state vanishes. We further 
	calculate the entanglement between the battery and charger to explore the 
	effect of entanglement in extracting the amount of energy from the QB. 
	Since the total system is always in a pure state, we use linear entropy to 
	quantify the entanglement between battery and charger {taken as two }
	subsystems\cite{Mitra_2024,MitraSrivastava_2024}. The linear entropy of the 
	QB is defined as
	\begin{equation}\label{eq:entropy_expression}
		S_L \left(\rho_b^n\right)=1- \tr\left(\rho_b^{n}\right)^2= 
		2\left(\lvert {(F_n)}_{21}\rvert^2 -\lvert {(F_n)}_{21}\rvert^4\right)
	\end{equation}
	As can be seen from Fig.~\ref{fig:energy_storage}, by tuning the parameters 
	of the Hamiltonian in Eq.~\ref{eq:tot_ham}, under the periodic 
	delta{-kicks}, it is possible to extract all 
	possible energy that is stored in the battery, which is an important 
	criterion for the best performance of the QB. For the special case where 
	$\kappa \tau=\frac{\pi}{2}$, the reduced density matrix of the battery is 
	simply $|0\rangle$ {or $\kt{1}$ depending on} the number 
	of kicks {being} even {or} odd. Therefore, it is possible to 
	realize optimal work extraction without producing any amount of 
	entanglement between the battery and charger. However, this is a very 
	special case where the coupled system behaves {like} two non-interacting 
	subsystems. As it is 
	shown in Fig.~\ref{fig:energy_storage}, it is also possible to 
	simultaneously realize optimal energy storage and optimal work extraction 
	from the QB for the other parameters as well when there is a finite amount 
	of entanglement between the battery and charger. 
	Specifically, in the limit $\kappa \gg \delta$, it is possible to realize 
	both optimal 
	energy storage and extraction. This {is better 
	than} central spin model{ where }optimal 
	energy storage { was }realized only for a set of particular 
	initial state of the charger \cite{Yang_2025}. In  
	Fig.~\ref{fig:energy_storage}, by choosing the 
	parameter $\kappa \gg \delta$, we numerically calculate the stored energy, 
	ergotropy, and linear entropy and then compare the numerical values with 
	the analytical expressions obtained in Eq.~\ref{eq:stored_eng_expression}, 
	Eq.~\ref{eq:ergotropy_expression} and Eq.~\ref{eq:entropy_expression}, 
	respectively. Interestingly, all the analytical expressions derived for 
	$h=0$ are in excellent agreement with the numerical values calculated for 
	small yet finite $h (\sim 0.2)$. As can be seen from the figure, optimal 
	energy can be stored and, at the same time, can be extracted without any 
	waste of energy at regular intervals of time. At this point, the linear 
	entropy between the battery and charger is minimized {to a non-zero value}. 
	The optimal energy storage and 
	work extraction are not confined to a 
	single value of $\tau$; instead, they are realized over finite intervals of 
	$\tau$, specifically within the ranges $0.02<\tau<0.16$ and 
	$0.18<\tau<0.35$ for fixed $\kappa=6$. For the same choice of parameter 
	regime, optimality in energy storage and extraction can be realized for a 
	higher number of batteries too. 
	
	In this limiting case when $h/J \ll 1$, we do not find any quantum 
	advantage since the maximum amount of entanglement between the battery and 
	charger subsystems does not scale with the number of batteries, the maximum 
	value is always $1/2$ irrespective of the number of batteries. As a result, 
	the average charging power at the charging time scales only linearly with 
	the number of batteries. Let us recall at this point that any superlinear 
	scaling of average charging power with the number of QBs is dubbed as the 
	quantum advantage \cite{Campaioli_2017}. Therefore, in this parameter 
	regime, quantum advantage in the charging process can not be achieved. 
	Furthermore, neither the energy storage nor ergotropy saturates around a 
	fixed value; rather, they show an oscillating behavior. A good QB should 
	provide optimal energy storage and extraction and, at the same time, should 
	be able to provide stable energy storage without much fluctuation and 
	produce a quantum advantage in the charging process. In the parameter 
	regime $J>>h$, the first criterion of producing a good QB is already 
	achieved; however, the other two criteria are yet to be satisfied. In the 
	next section, we investigate whether the remaining two criteria can be 
	achieved in the sunburst quantum Ising battery under periodic 
	delta{-kicks}  by introducing a finite transverse field 
	$h$, such that $J \sim h$.
	\section{Stable energy storage and quantum advantage}\label{sec:stability_advantage}
	{After achieving the near optimal energy storage and complete 
	extraction of energy for work, we focus on stability of the}  energy 
	storage\cite{Rosa_2020,Santos,Gherardini,Yao_Rydberg}  as well as 
	super-linear scaling of the average charging power, i.e, quantum advantage 
	{in this section}. For other QB models, which mostly lie in the 
	integrable regime, even though achieving super-linear scaling of charging 
	power is possible, stable energy storage cannot be ensured 
	\cite{andolina_25,Yang_2025,Grazi}. As  seen from the 
	{RMT studies } of the sunburst quantum 
	Ising battery under periodic delta{-kicks}, 
	the model {can be tuned from integrable to } 
	quantum chaotic regime. Therefore, we expect of achieving a stable energy 
	storage from this battery {by suitably choosing the parameters of the 
	model to put it in chaotic regime}.
	\begin{figure}
		\centering
		\includegraphics{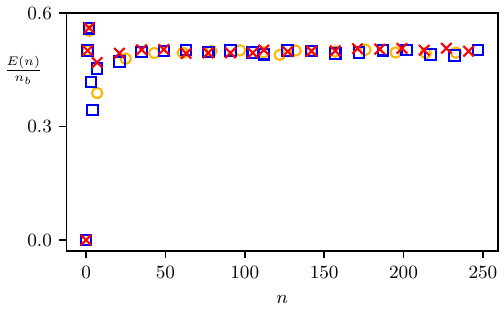}
		\caption{The dynamics of $\frac{E(n)}{n_b}$ is shown for $n_b=4$(dark 
		yellow), $5$(blue) and $6$(red). The total system sizes 
		are chosen as $L+n_b=13$ and the parameters 
		are chosen as $J,h,\kappa,\delta=1$ and $\tau=\pi/4$.}
		\label{fig:stable_energy}
	\end{figure}
	{In the limit $J\simeq h$, } we observe that the energy storage tends to 
	saturate 
	around a fixed value with negligible fluctuation (see 
	Fig.~\ref{fig:stable_energy}). We use the standard deviation of time-series 
	of $E(n)/n_b$  normalized by its mean value as the measure of temporal 
	fluctuation of stored energy.
	The time-series starts from the charging time to $T=250\tau$.
	For $n_b=1$ to $n_b=6$ with fixed $L+n_b=13$, the values are found to be 
	$0.07,0.07,0.05,0.04,0.03,$ and $0.03$ respectively. Clearly, the 
	energy storage becomes progressively more stable with increasing number of 
	batteries. Even for a single battery, stability is very good.

  \color{black}


	\begin{figure}
		\centering
		\includegraphics{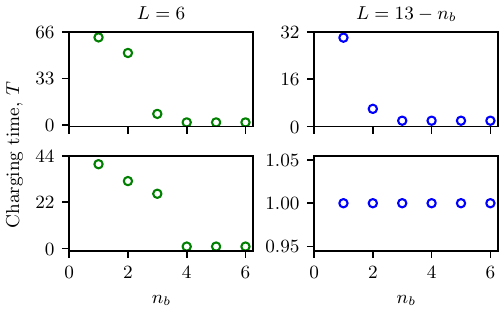}
		\caption{The charging time is plotted against $n_b$ in the left column 
		for fixed charger length $L=6$. The top-left curve corresponds to 
		$\tau=\pi/4$ while the bottom-left correspond to $\tau=\pi/4+0.1$. The 
		right column corresponds to scenario in which the total length of 
		charger and battery is kept constant, 13 to be precise. Like in the 
		left column case, top plot correspond to $\tau=\pi/4$ while bottom plot 
		corresponds $\tau=\pi/4+0.1$. Other parameters are chosen as 
		$J,h,\kappa,\delta=1$.}
		\label{fig:charging_time}
	\end{figure}
	{Next we 
	turn to study the possibility of obtaining }quantum advantage in the 
	charging process. For a classical battery, where no entanglement exists 
	between the battery and charger, the stored energy scales linearly with the 
	number of batteries, while the charging time remains independent of it. 
	Consequently, the charging power also scales linearly with the number of 
	batteries. In the presence of finite entanglement between the battery and 
	charger subsystems, the charging time can decrease with an increasing 
	number of batteries. As the stored energy still scales linearly with the 
	number of batteries, this reduction in charging time leads to a 
	super-linear scaling of the average charging power. In 
	Fig.~\ref{fig:charging_time}, we plot the charging time as a function of 
	the number of batteries. We observe that the charging power decreases as 
	$n_b$ increases from 1 to 4, and then saturates for $n_b > 4$, indicating 
	the absence of quantum advantage beyond this point. However, for $n_b > 4$, 
	the stored energy reaches its maximum immediately after two successive 
	kicks. This implies that for larger $n_b$, the battery subsystem can reach 
	the excited eigenstate more rapidly. We further ask whether, in the 
	sunburst quantum Ising battery, maximum energy storage can be achieved 
	immediately after the first kick, which would correspond to 
	{minimum charging time} of the QB system. As 
	shown in Fig.~\ref{fig:charging_time}, this is indeed possible when the 
	interval between successive kicks is chosen as $\tau \in [\pi/4 + 0.1, 
	\pi/4 + 0.2]$. Moreover, for this choice of parameter, the quantum 
	advantage for $n_b \leq 4$ remains intact. 

	\section{Nature of advantage}\label{sec:adv_nature}
	Having established the existence of quantum advantage in the previous 
	section, we now investigate whether this advantage arises purely from the 
	quantum correlations between the battery spins. In earlier literature on 
	quantum batteries, the advantage has been classified into \emph{genuine 
	quantum advantage} and \emph{classical advantage}, based on the source of 
	the super-extensivity in the scaling of charging power with the number of 
	batteries 
	\cite{Julia_2020,Campaioli_2017,Ferraro_2018,Rossini_2020,large_battery_Gao,Yang_2025,
	 andolina_25,Grazi}. If the super-extensivity arises due to the 
	multipartite nature of the entanglement between the battery spins, the 
	advantage is referred to as a genuine quantum advantage 
	\cite{Rossini_2020,andolina_25,puri_2024}. On the other hand, if it arises 
	from the super-extensive scaling of the speed of evolution in energy 
	eigenspace, the advantage is considered classical in nature. Based on the 
	geometric framework, the average charging power $P(T)$ is bounded by 
	\cite{Julia_2020},
	\begin{equation}\label{eq:bound_power}
	P(T) \leq 2\sqrt{\langle {\Delta H_b^2 (t)}\rangle_{t=T} \langle {\Delta H^2 (t)}\rangle_{t=T}}\equiv P_{\rm bo}(T),
	\end{equation} 
	where $\langle {\Delta H_b^2 (t)}\rangle_{t=T}$ ($\langle {\Delta H^2 (t)}\rangle_{t=T}$) denotes the time-average variance of the battery (total) Hamiltonian, evaluated at the charging time $T=m\tau$ and defined as
	\begin{equation}
	\langle {\Delta H_b^2 (t)}\rangle_{t=T}=\frac{1}{T}\sum_{t=0}^{t=m \tau} \Delta H_b^2(t)|_{t=T}.
	\end{equation} 
	The time-average variance of the battery Hamiltonian can scale super-linearly with the number of batteries only in the presence of multipartite entanglement between the battery spins, whereas ${\Delta H ^2(t)}$ determines the speed of evolution in energy eigenspace \cite{Julia_2020}.
	
	\begin{figure}[H]
	    \centering
	    \includegraphics{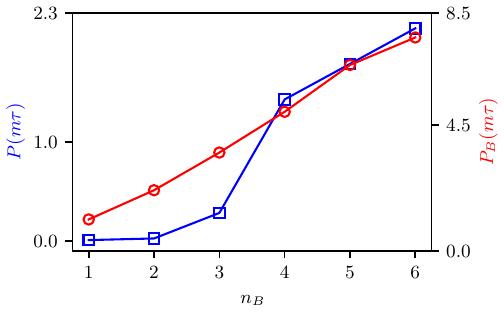}
	    \caption{Power from Eq:~\ref{eq:power_defn} (on the left $y$-axis)and 
	    the 
	    corresponding bound from Eq:~\ref{eq:bound_power} (on the right 
	    $y$-axis) 
	    have been plotted at the charging 
	    time ($m\tau$) with the $n_b$. A non-linear increase in power with 
	    increasing number of batteries is a clear signature of advantage over 
	    parallel classical charging scheme.}
	    \label{fig:p_bound}
	\end{figure}
	In Fig.~\ref{fig:p_bound}, we show that the scaling of the charging power with the number of batteries does not follow the scaling of $P_{\rm bo}(T)$. The charging power exhibits super-linear scaling with $n_b$ for $n_b \leq 4$, whereas $P_{\rm bo}(T)$ always scales linearly with $n_b$. This suggests that the bound in the average charging power derived in Eq.~\ref{eq:bound_power} is not tight enough to identify the origin of the super-linear scaling of the charging power \cite{andolina_25}. Therefore, to determine whether the super-linear scaling of the charging power arises from quantum correlations—and thus a genuine quantum advantage—or from some other source which are not directly linked to the generation of multipartite entanglement between the battery spins (and hence a classical advantage), we calculate the quantum Fisher information, which quantifies the multipartite entanglement between the battery spins\cite{Hong_qfi_21,Dell_Anna_qfi,Mathew_qfi,Yang_2025}.

	We consider a generic mixed state of $n_b$ qubits in its spectral decomposition
    \begin{equation}
        \rho= \sum_{i=1}^{2^{n_b}} \lambda_i \ket{\lambda_i}\bra{\lambda_i},
    \end{equation}
    where $\ket{\lambda_i}$ are eigenvector of $\rho$ with corresponding eigenvalue $\lambda_i$. For the state $\rho$, the quantum Fisher information is defined as \cite{Y_Su,Hyllus,toth}
		\begin{equation}
		\text{F}_Q[\rho,H]= 2\sum_{i=1}^{2^{n_b}} \sum_{j=1}^{2^{n_b}} \frac{(\lambda_i -\lambda_j)^2}{(\lambda_i+ \lambda_j)}  \bra{\lambda_i}\hat{S^\alpha}\ket{\lambda_j}\bra{\lambda_j}\hat{S^{\alpha^\prime}}\ket{\lambda_i},
	\end{equation}
	where $\hat{S^\alpha}=\frac{1}{2}\hat{\sigma^\alpha}$ is the spin operator and $\alpha=\{x,y,z\}$. For the reduced time evolved state of the battery, $\rho_b^n$, the multipartite entanglement between battery qubits is given by 
		\begin{equation}
		F[\rho(t), \vec{n}.\vec{S}]\equiv \max_{\lvert \vec{n}\rvert=1}\vec{n}^T\Gamma \vec{n}\equiv  \lambda_{max}(\Gamma),
	\end{equation}
	where $\vec{n}$ is a unit vector on the Bloch sphere. The matrix $\Gamma$ is introduced for maximization of $F[\rho(t), \vec{n}.\vec{S}]$ over measurement directions and its elements are defined as
	\begin{equation}
		\Gamma_{\alpha, \alpha^\prime}=2\sum_{i,j} \frac{(\lambda_i -\lambda_j)^2}{(\lambda_i+ \lambda_j)}  \bra{\lambda_i}\hat{\text{S}^\alpha}\ket{\lambda_j}\bra{\lambda_j}\hat{\text{S}^{\alpha^\prime}}\ket{\lambda_i}.
	\end{equation}
	Here, $\lambda_{max}(\Gamma)$ is the largest eigenvalue of $\Gamma$ matrix. A multipartite entanglement witness is then obtained by the following condition:
    \begin{equation}
        \lambda_{max}(\Gamma)>n_b.
    \end{equation}

\begin{table}[h]
\begin{tabular}{|c|c|c|c|c|}
\hline
$n_b$ &
$L=6$, $\tau=\pi/4$ &
$L=6$, $\tau=\pi/4+0.1$ &
$L=13-n_b$, $\tau=\pi/4$ &
$L=13-n_b$, $\tau=\pi/4+0.1$ \\
\hline
2 & 0.287 & 0.047 & 0.219 & 0.191 \\
\hline
3 & 0.205 & 0.082 & 0.187 & 0.410 \\
\hline
4 & 1.395 & 0.278 & 0.262 & 1.202 \\
\hline
5 & 0.664 & 2.942 & 1.220 & 0.587 \\
\hline
6 & 2.260 & 4.655 & 1.099 & 1.531 \\
\hline
\end{tabular}
\caption{Maximum eigenvalue $\lambda_{\max}(\Gamma)$ for various $n_b$, $L$, 
and $\tau$ values.}
\label{tab:QFI}
\end{table}

In Tab.~\ref{tab:QFI}, we summarize the quantum Fisher information computed 
under two settings. First, we fix the number of Ising sites at $L=6$ and vary 
the number of battery qubits $n_b=2,...,6$ for $\tau=\pi/4$ and 
$\tau=\pi/4+0.1$. Second, we fix the total system size $L+n_b=13$ and again 
vary $n_b=2,...,6$ using the same two choices of $\tau$. In every cases, we 
find that $\lambda_{max}(\Gamma)<n_b$, indicating the absence of multipartite 
entanglement between the battery qubits. This allows us to conclude that, for 
the sunburst quantum Ising battery driven by periodic delta-kicks, the observed 
advantage in the charging process does not originate from multipartite 
entanglement and therefore does not constitute a genuine quantum advantage. 
Instead, the advantage  {is classical  
}similar to what is observed in the central spin battery\cite{Yang_2025}.

\section{Conclusion}\label{sec:conc}
In this paper, we {have studied }the performance of the 
sunburst quantum Ising battery under periodic delta{-kicks}. In the case of the 
interaction quench protocol, where the external 
control parameter $\lambda(t)$ is a step function, the charging time does not 
depend on the number of batteries $n_b$. In contrast, under periodic 
delta{-kicks} the charging time decreases with increasing 
$n_b$ for $n_b \leq 4$. This indicates a super-linear scaling of the charging 
power with $n_b$, thus signalling the presence of quantum advantage in the 
charging process. For $n_b>4$, the stored energy reaches its maximum after the 
very first kick, implying {near }instantaneous charging in the presence 
of a large number of QBs. We further show that this quantum advantage does not 
originate from multipartite entanglement among the battery qubits and therefore 
is {classical in nature}. In 
contrast to commonly studied QB models that typically lie in the integrable 
regime, we show by analyzing the statistical properties within the framework of 
random matrix theory that the sunburst quantum Ising battery under periodic 
delta{-kicks} belongs to the quantum chaotic 
regime. Consequently, this battery can simultaneously provide stable energy 
storage along with quantum advantage. Although quantum advantage has been 
reported in many quantum batteries based on integrable models 
\cite{andolina_25, Yang_2025,Le_2018}, these systems cannot naturally achieve 
stable energy storage due to integrability. Moreover, unlike the 
Sachdev–Ye–Kitaev battery\cite{Rossini_2020, Rosa_2020}, the present model 
allows both quantum advantage and stable energy storage within the same 
parameter regime keeping only two body interaction. In the near-integrable 
regime, both optimal energy storage 
and energy extraction are possible irrespective of the charger’s initial state. 
This feature is absent in the central spin battery, where optimization can only 
be achieved for specific initial states of the charger\cite{Yang_2025}. In 
summary, the sunburst quantum Ising battery uniquely combines all three desired 
features of an efficient quantum battery—optimal energy storage and extraction, 
stable energy storage, and advantage in the charging process. To the best of 
our knowledge, these three features have not been found together in any other 
quantum battery model.

\section*{Funding}
Open access funding provided by Department of Atomic Energy.

\section*{Author contributions}
All authors contributed equally to the analysis, interpretation of the results 
and the preparation of the manuscript.
	\bibliographystyle{unsrtnat}
	\bibliography{references_kick}
	
\end{document}